\documentclass[11pt]{article}

\usepackage[top=1.3in, bottom=1.3in, left=1.25in, right=1.25in]{geometry}

\usepackage{fancyhdr} 
\usepackage{lastpage} 
\usepackage{extramarks} 
\usepackage[usenames,dvipsnames]{color} 
\usepackage{graphicx} 
\usepackage{listings} 
\usepackage{courier} 
\usepackage{amsmath}
\usepackage{amsfonts}
\usepackage{mathtools}
\usepackage{subfigure}
\usepackage{enumitem}
\usepackage{booktabs}
\usepackage{bm}
\usepackage{url}
\usepackage{ucs}
\usepackage{graphicx}
\usepackage[utf8x]{inputenc}
\usepackage{caption}
\usepackage{verbatim}
\usepackage{footmisc}

\usepackage{algorithm, algorithmic}

\graphicspath{{plots/}}

\definecolor{MyDarkGreen}{rgb}{0.0,0.4,0.0} 
\begin{document}

\bibliographystyle{unsrt}

\title{Optimization on the Surface of the (Hyper)-Sphere}
\author{
Parameswaran Raman\footnotemark[1]\\
       {University of California, Santa Cruz}\\
       {params@ucsc.edu}
\and
Jiasen Yang\\
        {Purdue University}\\
        {jiaseny@gmail.com}
}

\footnotetext[1]{This work was done while the author was a first year Ph.D. student at Purdue University (2013-2014).}
\date{} 

\maketitle

\begin{abstract}
Thomson problem is a classical problem in physics to study how $n$ number of charged particles distribute themselves on the surface of a sphere of $k$ dimensions. When $k=2$, i.e. a 2-sphere (a circle), the particles appear at equally spaced points. Such a configuration can be computed analytically. However, for higher dimensions such as $k \ge 3$, i.e. the case of 3-sphere (standard sphere), there is not much that is understood analytically. Finding global minimum of the problem under these settings is particularly tough since the optimization problem becomes increasingly computationally intensive with larger values of $k$ and $n$. In this work, we explore a wide variety of numerical optimization methods to solve the Thomson problem. In our empirical study, we find stochastic gradient based methods (SGD) to be a compelling choice for this problem as it scales well with the number of points.
\end{abstract}

\thispagestyle{empty}


\setcounter{page}{1}

\section{Introduction}

\subsection{The Thomson Problem}

The \emph{Thomson problem} was posed by physicist J. J. Thomson to determine the minimum electrostatic potential energy configuration of $n$ electrons on the surface of a unit sphere that repel each other with a force given by \emph{Coulomb's law} \cite{WikiThomson}.

It is clear that for the case of $n=2$, the optimal configuration consists of electrons at antipodal points.
For $n=3$, electrons reside at the vertices of an equilateral triangle about a great circle.
Minimum energy configurations have also been rigorously identified for the cases of $n = 4, 5, 6, 12$ \cite{WikiThomson}.
However, no analytical solutions have been found in more general cases.
This provides a good opportunity for us to study the problem using methods of numerical optimization.

\subsection{Mathematical Formulation}

In this project, our goal is to study the following  modification of the Thomson problem:
\begin{equation}\label{eq:Thomson1}
\begin{aligned}
& \text{minimize}
& & \sum_{i=1}^n\sum_{j=1}^{i-1}\frac{1}{\|\bm{x}_i-\bm{x}_j\|_2^2} \\
& \text{subject to}
& & \bm{x}_i\in\mathbb{R}^k,\ \|\bm{x}_i\|_2 = 1,\ 1 \le i \le n.
\end{aligned}
\end{equation}
where each $\bm{x}_i$ represents the coordinates of a point lying on the $k$-dimensional hyper-sphere.

For convenience, we can stack all the $\bm{x}_i$'s into a $k\times n$ matrix denoted by
$$ \bm{X} = \left[\begin{array}{cccc} \bm{x}_1 & \bm{x}_2 & \cdots & \bm{x}_n \end{array}\right], $$
and rewrite the optimization problem in the form
\begin{equation}\label{eq:Thomson2}
\begin{aligned}
& \text{minimize}
& & f(\bm{X}) := \sum_{i=1}^n\sum_{j=1}^{i-1}\frac{1}{\|\bm{x}_i-\bm{x}_j\|_2^2} \\
& \text{subject to}
& & \bm{X}\in\mathbb{R}^{k\times n},\ \text{diag}\{\bm{X}\bm{X}^\mathsf{T}\} = \bm{e},
\end{aligned}
\end{equation}
where $\bm{e}=[1,1,\cdots,1]^\mathsf{T}\in\mathbb{R}^k$.
In this report, our discussion will primarily focus on the case of $k=3$ since the solutions are easy to visualize. However, most of the methods we discuss can be directly applied to cases where $k$ is moderately large as well.

The remainder of this report is organized in the following manner.
In Section \ref{sec:OptMethods}, we apply a sequence of optimization methods to study Problem \eqref{eq:Thomson2}.
Apart from applying well-known optimization methods including the penalty method, the augmented Lagrangian method, the interior-point method, stochastic gradient descent, and the Nelder-Mead method, we also discuss unconstrained optimization via spherical coordinates and creatively introduce a \emph{Coulomb force method} to study the problem.
We provide visualizations of the obtained solutions and compare their accuracy using the converged values of the objective function.
In Section \ref{sec:Explore}, we explore more creative ideas to study the problem.
In particular, we develop a convex reformulation of the constraint, and briefly review the \emph{spherical packing problem}.



\section{Optimization Methods}\label{sec:OptMethods}


\subsection{Unconstrained Optimization via Spherical Coordinates}\label{sec:Sphere}

\begin{figure}[!htb]
  \centering
  \includegraphics[width=0.3\columnwidth]{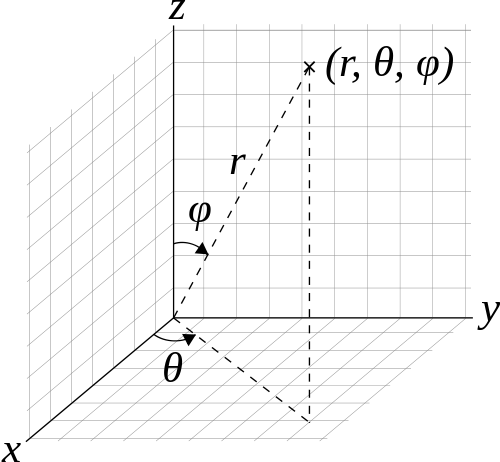}\\
  \caption{Illustration of the spherical coordinates \cite{WikiSphere}.}
  \label{fig:Spherical}
\end{figure}

For the case of $k=3$, problem \eqref{eq:Thomson2} becomes minimization of the objective function on the surface of the unit sphere in $\mathbb{R}^3$.
In this case, a natural way of converting the problem into an unconstrained optimization problem is to denote the points using the spherical coordinates
\begin{equation*}
\begin{dcases}
x = \sin\phi\,\cos\theta; \\
y = \sin\phi\,\sin\theta; \\
z = \cos\phi,
\end{dcases}
\end{equation*}
where $\phi\in[0,\pi]$ and $\theta\in[0,2\pi]$. Figure \ref{fig:Spherical} provides an illustration of the angles $\phi$ and $\theta$. 

Under the spherical coordinates, problem \eqref{eq:Thomson2} becomes the unconstrained problem
\begin{equation}\label{eq:Sphere}
\text{minimize}\quad f(\phi, \theta) := -\sum_{i=1}^n\sum_{j=1}^{i-1}\frac{1}{2[\sin\phi_i\sin\phi_j\cos(\theta_i-\theta_j) + \cos\phi_i\cos\phi_j-1]}.
\end{equation}
Differentiating with respect to $\phi_i$ and $\theta_i$ yields the gradients
\begin{align}
\frac{\partial f}{\partial \phi_i} & = \sum_{j\neq i}\frac{\cos\phi_i\sin\phi_j\cos(\theta_i-\theta_j)-\sin\phi_i\cos\phi_j}{2[\sin\phi_i\sin\phi_j\cos(\theta_i-\theta_j) + \cos\phi_i\cos\phi_j-1]^2}, \\
\frac{\partial f}{\partial \theta_i} & = -\sum_{j\neq i}\frac{\sin\phi_i\sin\phi_j\sin(\theta_i-\theta_j)}{2[\sin\phi_i\sin\phi_j\cos(\theta_i-\theta_j) + \cos\phi_i\cos\phi_j-1]^2}.
\end{align} 

We check the correctness of our gradient computations using the forward finite differences method implemented in the \texttt{gradientcheck} function contained in the Poblano toolbox. The MATLAB wrapper code and output is shown as follows:

\begin{lstlisting}
% Gradient Check for spherical_main.m
n = 2;
phi = 2*pi*rand(1,n);
theta = pi*rand(1,n);
out = gradientcheck(@(x) spherical_obj(x, n), [phi theta]',
		'DifferenceType','forward');
out
out.G
out.GFD
\end{lstlisting}

\begin{verbatim}
>> gradcheck
out = 
                    G: [4x1 double]
                  GFD: [4x1 double]
              MaxDiff: -7.6871e-06
           MaxDiffInd: 1
    NormGradientDiffs: 9.4885e-06
        GradientDiffs: [4x1 double]
               Params: [1x1 struct]

ans =
 -148.9590
   55.3360
  -37.4119
   37.4119

ans =
 -148.9590
   55.3360
  -37.4119
   37.4119
\end{verbatim}

\noindent From the above output, we verify that the numerical gradients matches our analytically computed gradients with high precision. 

\begin{figure}[!htb]
  \centering
  \subfigure[$n=2$]{
    \includegraphics[width=0.4\columnwidth]{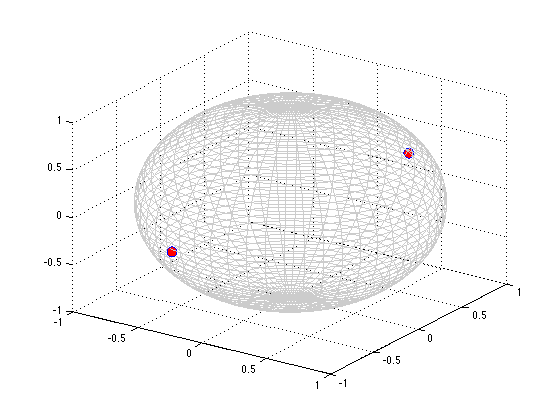}}
  \subfigure[$n=3$]{
    \includegraphics[width=0.4\columnwidth]{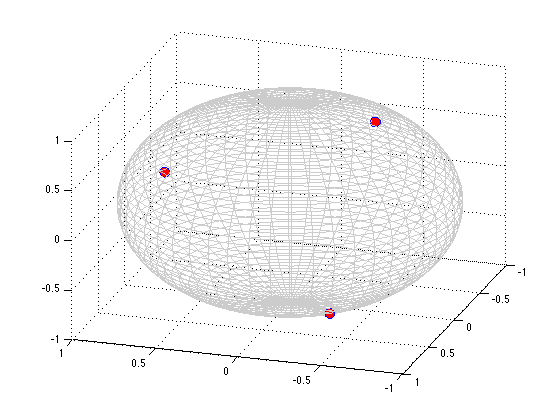}}
  \subfigure[$n=4$]{
    \includegraphics[width=0.4\columnwidth]{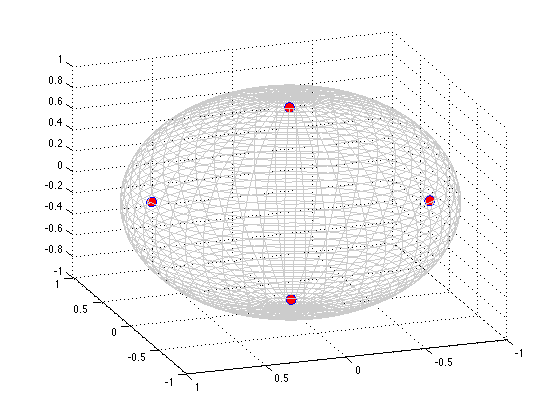}}
  \subfigure[$n=10$]{
    \includegraphics[width=0.4\columnwidth]{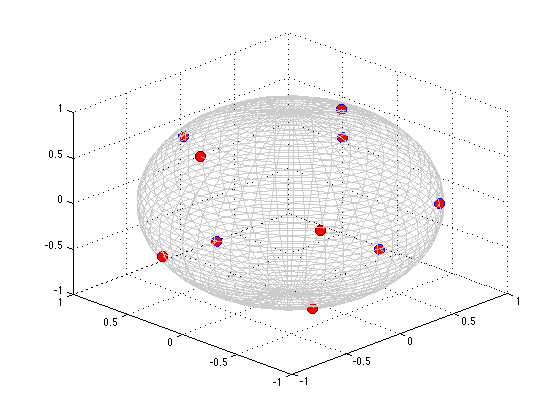}}
  \caption{Visualization of the solutions for the unconstrained spherical coordinates method.}
  \label{fig:SphereResults}
\end{figure}

To solve the optimization problem, we notice that evaluation of the objective function and its gradients takes $\mathcal{O}(n^2)$ computation, whereas evaluation of the Hessian would take $\mathcal{O}(n^3)$ computation.
We therefore utilize Quasi-Newton methods such as L-BFGS to solve the optimization problem.
The \texttt{lbfgs} function in Poblano is used to solve problem \eqref{eq:Sphere} for various values of $n$. Figure \ref{fig:SphereResults} visualizes the solutions for the cases of $n=2, 3, 4, 10$.

\subsection{Projected Gradient Descent}
The methods discussed beyond this section can be directly applied to cases where $k\ge3$. 
Differentiation of the objective function in \eqref{eq:Thomson2} yields the gradients
\begin{equation}
\frac{\partial f(\bm{X})}{\partial x_{ij}} = -2\sum_{l\neq i}\frac{x_{ij}-x_{lj}}{\|\bm{x}_i-\bm{x}_l\|_2^4}, \quad 1\le i,j\le n;
\end{equation}
which can be written more compactly as
\begin{equation}\label{eq:gradient}
\nabla_{\bm{x}_i} f(\bm{X}) = -2\sum_{l\neq i} \frac{\bm{x}_i-\bm{x}_l}{\|\bm{x}_i-\bm{x}_l\|_2^4}, \quad 1\le i\le n.
\end{equation}

There are many ways to reformulate the constraints in \eqref{eq:Thomson2} in order to convert it into an unconstrained problem, as we shall demonstrate in subsequent sections.
However, let us start by considering a simple and intuitive algorithm, in which we perform gradient descent on the original objective function in \eqref{eq:Thomson2}, and after each iteration we project the updated points $\bm{x}_1,\cdots\bm{x}_n$ back onto the surface of the (hyper)-sphere (which corresponds to normalizing the columns of $\bm{X}$).
Clearly, the convergence of this algorithm could not be guaranteed; yet this algorithm appears to work empirically under careful initializations.


\subsection{Penalty Method}

In this section, we convert problem \eqref{eq:Thomson2} into an unconstrained problem by relaxing the constraint into the objective function using a regularization parameter $\lambda$ which could be tuned by cross validation.

The penalized objective function takes the form
\begin{equation}\label{eq:PenaltyObj}
f(\bm{X}) := \sum_{i=1}^n\sum_{j=1}^{i-1}\frac{1}{\|\bm{x}_i-\bm{x}_j\|_2^2} + \frac{\lambda}{2} \sum_{i=1}^n \left(\|\bm{x}_i\|^2-1\right)^2.
\end{equation}
Here, we modified the constraint to $(\|\bm{x}_i\|^2-1)^2/2=0$ in order to avoid cancellation of positive/negative errors.
The gradients of the penalized objective function \eqref{eq:PenaltyObj} are given by
\begin{equation}
\nabla_{\bm{x}_i} f(\bm{X}) = -2\sum_{l\neq i} \frac{\bm{x}_i-\bm{x}_l}{\|\bm{x}_i-\bm{x}_l\|_2^4} + \lambda\left(\|\bm{x}_i\|^2-1\right)\bm{x}_i, \quad 1\le i\le n.
\end{equation}

The intuition behind the penalty method is that as $\lambda\to\infty$, minimizing the objective function \eqref{eq:PenaltyObj} should be equivalent to solving the original constrained optimization problem \eqref{eq:Thomson2}. In other words, for large values of $\lambda$, minimizers of \eqref{eq:PenaltyObj} should approximately lie on the surface of the unit sphere; this might not be the case for small values of $\lambda$. However, since accurately solving the penalized problem with a large value of $\lambda$ give rise to computational difficulties, we instead start by setting the value of $\lambda$ to be quite small, say $\lambda=1$, and then solve the penalized problem; after which we gradually increase the value of $\lambda$ in small increments, solving each more heavily penalized problem with $\bm{X}$ initialized as the solution to the previous problem. Solving this sequence of penalized problems would eventually provide us with a solution with high precision.

\begin{figure}[!htb]
  \centering
  \subfigure[$\lambda=1$]{
    \includegraphics[width=0.45\columnwidth]{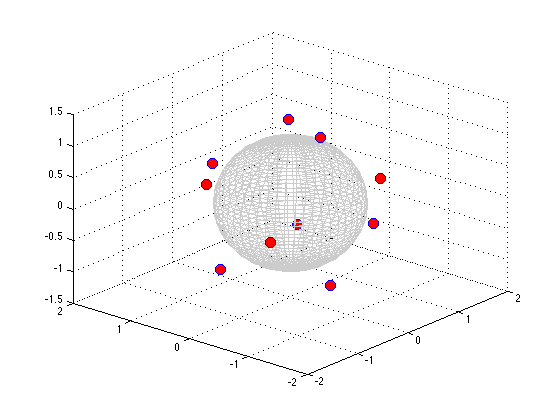}}
  \subfigure[$\lambda=100$]{
    \includegraphics[width=0.45\columnwidth]{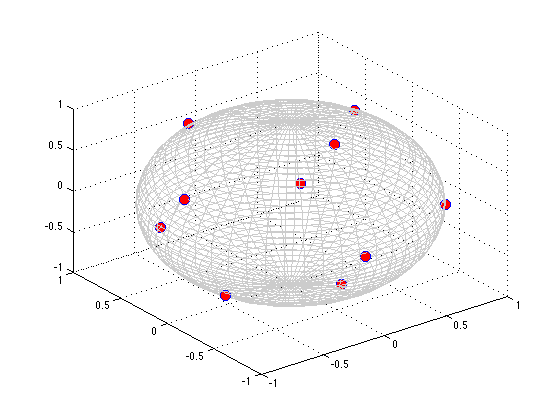}}
  \caption{Visualization of the solutions for the penalty method ($n=10$).}
  \label{fig:PenaltyResults}
\end{figure}

Figure \ref{fig:PenaltyResults} illustrates the intuition we described. In the figure, we observe that for the penalized problem with $\lambda=1$, the constraint is poorly enforced in that none of the points lie on the surface of the unit sphere. However, for the penalized problem with $\lambda=100$, the constraint appears to be very well approximated.

As before, we verify the correctness of our gradient computations using the forward finite differences method as follows:

\begin{lstlisting}
% Gradient Check for penalty_main.m
N = 2;
p = 3;
X = 2*rand(n, p)-1;
x = X; x = x(:);
out = gradientcheck(@(x) penalty_obj(x, n, p, 1), x,
		'DifferenceType', 'forward');
out
out.G
out.GFD
\end{lstlisting}

\begin{verbatim}
>> gradcheck
out = 
                    G: [6x1 double]
                  GFD: [6x1 double]
              MaxDiff: -7.1876e-07
           MaxDiffInd: 4
    NormGradientDiffs: 1.0145e-06
        GradientDiffs: [6x1 double]
               Params: [1x1 struct]

ans =
   -5.0305
    4.4829
  -19.0028
   19.1971
   -1.1199
    1.2200

ans =
   -5.0305
    4.4829
  -19.0028
   19.1971
   -1.1199
    1.2200
\end{verbatim}

\noindent Having verified the gradients, we utilize the \texttt{lbfgs} function in Poblano to minimize the objective function \eqref{eq:PenaltyObj} for various values of $n$. Some converged values are listed in Table \ref{tb:ObjReg}.

\begin{table}[ht]
  \centering
  \begin{tabular}{rcccc}
	\toprule
	$n$ & 10 & 20 & 30 & 40 \\
	\midrule
	$f(\bm{X}^*)$ & 24.7424 & 129.9554 & 342.4396 & 659.5119 \\
	\bottomrule
  \end{tabular}
  \caption{Converged values of the objective function for the penalty method.}
  \label{tb:ObjReg}
\end{table}

\subsection{Augmented Lagrangian Method}

The augmented Lagrangian method can be viewed as an extension of the penalty method, and the objective function takes on a similar form:
\begin{equation}\label{eq:RegObj}
f(\bm{X}) := \sum_{i=1}^n\sum_{j=1}^{i-1}\frac{1}{w\|\bm{x}_i-\bm{x}_j\|_2^2} + \frac{\lambda}{2} \sum_{i=1}^n \left(\|\bm{x}_i\|^2-1\right)^2 - \sum_{i=1}^n \mu_i\left(\|\bm{x}_i\|^2-1\right),
\end{equation}
with its gradients given by
\begin{equation}\label{eq:gradient}
\nabla_{\bm{x}_i} f(\bm{X}) = -2\sum_{l\neq i} \frac{\bm{x}_i-\bm{x}_l}{\|\bm{x}_i-\bm{x}_l\|_2^4} + 2\lambda\left(\|\bm{x}_i\|^2-1\right)\bm{x}_i - 2\mu_i\bm{x}_i, \quad 1\le i\le n.
\end{equation}

We note that in both the penalty method and the augmented Lagrangian method, since the constraint is only loosely approximated in the beginning, we do not need to enforce a very stringent tolerance level as the solutions are will only be serving as initial values for the next problem in the sequence. Therefore, computationally it would be more efficient to gradually decrease the tolerance setting as $\lambda$ increases, which would save us time in solving the initial problems.

As always, we verify the correctness of our gradient computations using the forward finite differences method as follows:
\begin{lstlisting}
% Gradient Check for augmented_lagrangian_main.m
N = 3;
p = 3;
X = 2*rand(n, p)-1;
for i = 1:n
    X(i,:) = X(i,:)/norm(X(i,:));
end
x = X; x = x(:);
lambda = 1;
mu = zeros(n,1);
out = gradientcheck(@(x) augmented_lagrangian_obj(x,n,p,lambda,mu), x,
		'DifferenceType', 'forward');
out
out.G
out.GFD
\end{lstlisting}

\begin{verbatim}
>> gradcheck
out = 
                    G: [9x1 double]
                  GFD: [9x1 double]
              MaxDiff: 4.6926e-08
           MaxDiffInd: 4
    NormGradientDiffs: 7.6902e-08
        GradientDiffs: [9x1 double]
               Params: [1x1 struct]

ans =
    1.1957
    1.4398
   -2.6354
   -1.0928
    1.8569
   -0.7641
   -0.7383
   -1.5598
    2.2981

ans =
    1.1957
    1.4398
   -2.6354
   -1.0928
    1.8569
   -0.7641
   -0.7383
   -1.5598
    2.2981
\end{verbatim}

\noindent Having verified the gradients, we utilize the \texttt{lbfgs} function in Poblano to minimize the objective function \eqref{eq:RegObj} for various values of $n$. Some converged values are listed in Table \ref{tb:ObjAugLag}.

\begin{table}[!htb]
  \centering
  \begin{tabular}{rcccc}
	\toprule
	$n$ & 10 & 20 & 30 & 40 \\
	\midrule
	$f(\bm{X}^*)$ & 24.7452 & 129.9907 & 337.3002 & 655.3411 \\
	\bottomrule
  \end{tabular}
  \caption{Converged values of the objective function for the augmented Lagrangian method.}\label{tb:ObjAugLag}
\end{table}

\subsection{Interior-Point Method}

Although the interior-point method is commonly known as an effective method for solving optimization problems with inequality constraints, we also applied it to solve problem \eqref{eq:Thomson2}. In particular, we utilize the MATLAB implementation of the interior-point method by calling the \texttt{fmincon} function.

We apply the interior-point method to minimize the objective function \eqref{eq:PenaltyObj} for various values of $n$, and some converged values are listed in Table \ref{tb:ObjIntPt}. We notice that the converged function values for the penalty method, the augmented Lagrangian method, and the interior-point method are quite close to each other.

\begin{table}[ht]
  \centering
  \begin{tabular}{rcccc}
	\toprule
	$n$ & 10 & 20 & 30 & 40 \\
	\midrule
	$f(\bm{X}^*)$ & 25.0678 & 132.8934 & 338.0972 & 640.3781 \\
	\bottomrule
  \end{tabular}
  \caption{Converged values of the objective function for the interior-point method.}\label{tb:ObjIntPt}
\end{table}

\begin{figure}[!htbp]
  \centering
  \subfigure[$n=10$]{
    \includegraphics[width=0.8\columnwidth]{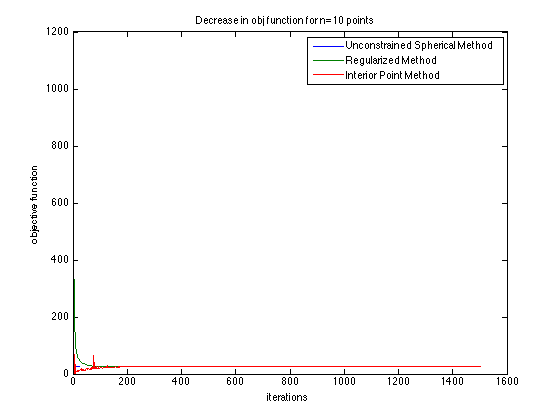}} \\
  \subfigure[$n=30$]{
    \includegraphics[width=0.8\columnwidth]{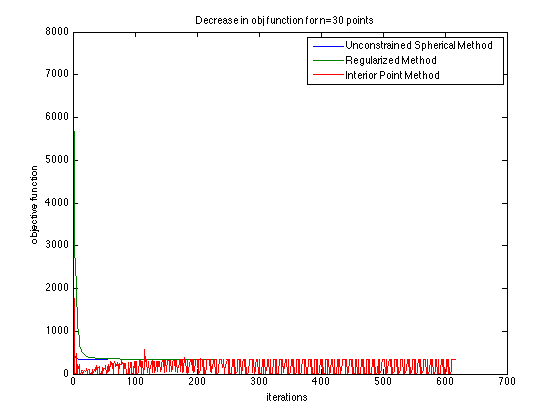}}
  \caption{Plots of the objective function value vs iteration number for the spherical method, the penalty (regularized) method, and the interior-point method.}
  \label{fig:ObjIter}
\end{figure}

\begin{figure}[htbp]
  \centering
  \subfigure[Spherical method]{
    \includegraphics[width=0.5\columnwidth]{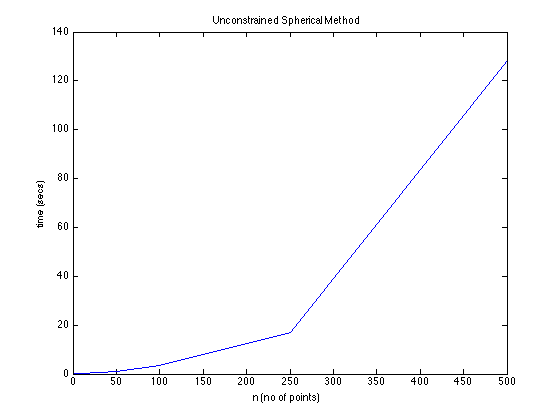}} \\
  \subfigure[Regularization method]{
    \includegraphics[width=0.5\columnwidth]{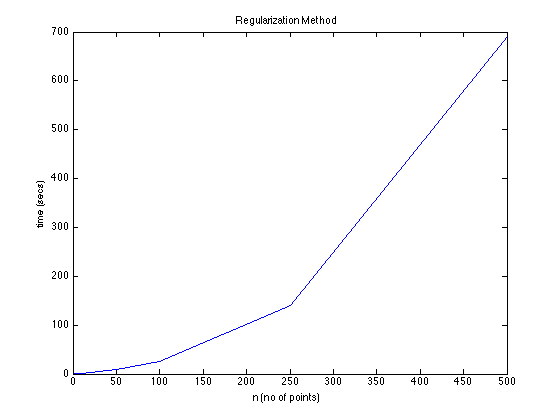}} \\
  \subfigure[Interior-point method]{
    \includegraphics[width=0.5\columnwidth]{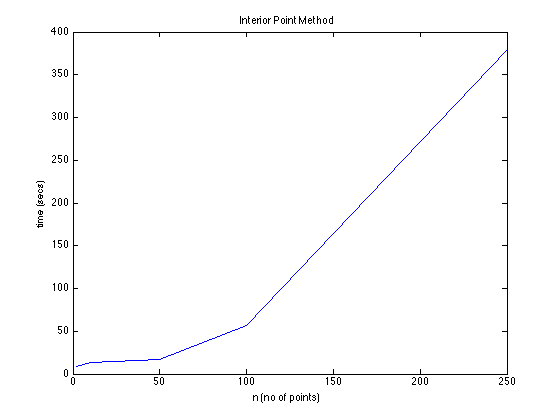}} \\
  \caption{Plots of the time until convergence for varying number of points.}
  \label{fig:TimeIter}
\end{figure}

Additionally, to compare the unconstrained spherical coordinates method, the penalty method, and the interior-point method, Figure \ref{fig:ObjIter} plots the value of the objective function against iteration number for the cases $n=10$ and $n=30$, while Figure \ref{fig:TimeIter} illustrates the time it takes until convergence for varying values of $n$.

\subsection{Stochastic Gradient Descent}

We remark that the evaluation of the gradient $\nabla_{\bm{x}_i}f(\bm{X})$ in equation \eqref{eq:gradient} requires a summation over $n-1$ points. Thus, if $n$ becomes very large, it can be quite costly to evaluate the gradient. In this case, instead of performing \emph{batch} gradient descent updates which takes $\mathcal{O}(n)$ computation, we could turn to utilize \emph{stochastic} gradient descent updates which takes $\mathcal{O}(1)$ computation instead. Even though the stochastic gradient descent algorithm would need a lot more iterations to converge, it can take much less running time in total as compared to batch gradient descent. In addition, stochastic gradient descent would also be an appealing algorithm if we wish to consider an \emph{online} version of the Thomson problem.

The objective function we use for stochastic gradient descent is the same as that of the penalty method, in which we relax the constraint using a regularization parameter $\lambda$.
\begin{equation}\label{eq:RegObjSGD}
f(\bm{X}) := \sum_{i=1}^n\sum_{j=1}^{i-1}\frac{1}{\|\bm{x}_i-\bm{x}_j\|_2^2} + \frac{\lambda}{2} \sum_{i=1}^n \left(\|\bm{x}_i\|^2-1\right)^2.
\end{equation}
The gradients of the penalized objective function \eqref{eq:RegObjSGD} are given by
\begin{equation}\label{eq:gradient}
\nabla_{\bm{x}_i} f(\bm{X}) = -2\sum_{l\neq i} \frac{\bm{x}_i-\bm{x}_l}{\|\bm{x}_i-\bm{x}_l\|_2^4} + \lambda\left(\|\bm{x}_i\|^2-1\right)\bm{x}_i, \quad 1\le i\le n.
\end{equation}
To derive the stochastic gradient descent algorithm, we introduce the notation
\begin{equation}\label{eq:gradient}
\nabla_{\bm{x}_i,\,\bm{x}_l} f(\bm{X}) = -2\frac{(\bm{x}_i-\bm{x}_l)}{\|\bm{x}_i-\bm{x}_l\|_2^4} + \frac{\lambda}{n-1}\left(\|\bm{x}_i\|^2-1\right)\bm{x}_i, \quad 1\le i\le n.
\end{equation}	
The stochastic gradient descent algorithm to minimize the above objective function is outlined in Algorithm \ref{alg:SGD}. Instead of summing over $n-1$ points to perform each gradient update, here we randomly sample a pair of points and perform stochastic updates only for that pair.

\begin{algorithm}
  \caption{Stochastic Gradient Descent}
  \label{alg1}
  \begin{algorithmic}\label{alg:SGD}
  \REQUIRE $n$ (\# of points), $p$ (dimension), $\lambda \text{ (regularization parameter)}$, $\gamma$ (step size), $iters$.
  \FOR{$t=1$ to $iters$}
  \STATE Randomly sample a pair of different points $\bm{x}_i,\bm{x}_l$ from the $n$ points.
  \STATE Update $\bm{x}_i,\bm{x}_l$ by scaling up the stochastic gradients:
    \begin{align*}
    \bm{x}_i & \leftarrow \bm{x}_i - \gamma (n-1) \nabla_{\bm{x}_i,\,\bm{x}_l} f(\bm{X}); \\
    \bm{x}_l & \leftarrow \bm{x}_l - \gamma (n-1) \nabla_{\bm{x}_i,\,\bm{x}_l} f(\bm{X});
    \end{align*}
\ENDFOR
 \end{algorithmic}
\end{algorithm}

\begin{figure}[ht]
  \centering
  \subfigure[$n=40$]{
    \includegraphics[width=0.45\columnwidth]{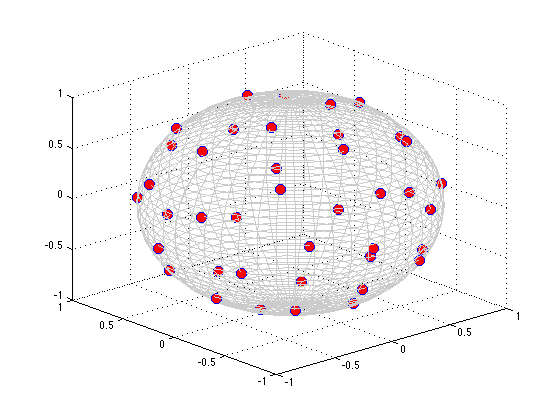}}  
  \subfigure[$n=100$]{
    \includegraphics[width=0.45\columnwidth]{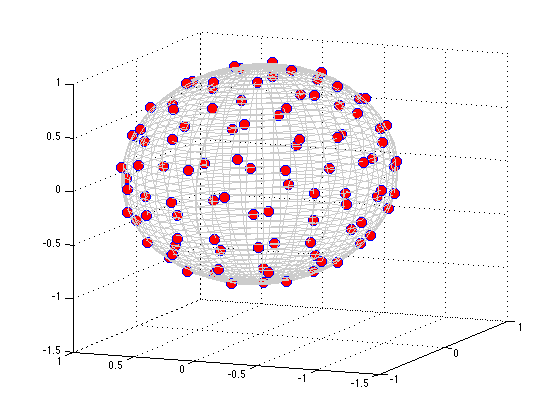}} 
  \caption{Plots of the solutions for stochastic gradient descent.}
  \label{fig:SGDPoints}
\end{figure}

\begin{figure}[ht]
  \centering
  \subfigure[$n=40$]{
    \includegraphics[width=0.45\columnwidth]{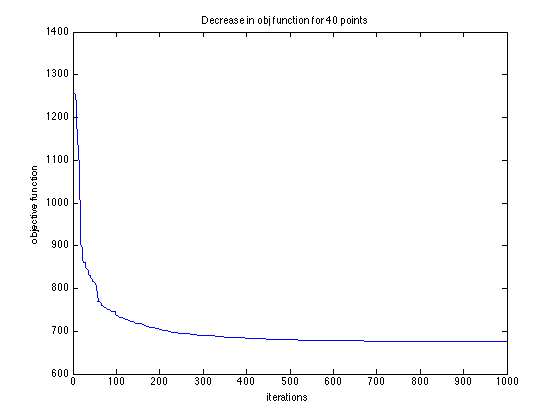}}
  \subfigure[$n=100$]{
    \includegraphics[width=0.45\columnwidth]{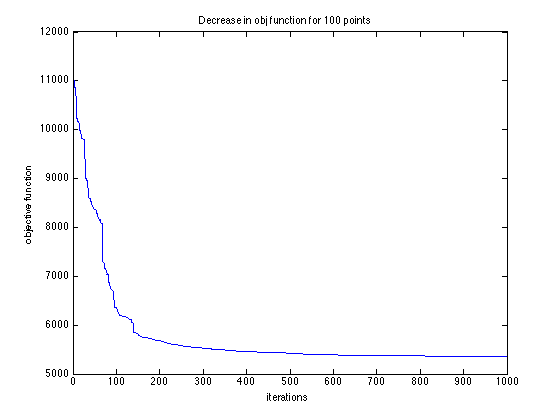}}
  \caption{Plots of the objective function value vs iteration for stochastic gradient descent.}
  \label{fig:SGDObj}
\end{figure}

Our experiments show that the algorithm converges to a local optima after carefully tuning the parameters. Compared to previous methods, stochastic gradient descent clearly wins out in terms of execution time when the number of data points $n$ is large. The solutions obtained for $n=40$ and $n=100$ are shown in Figure \ref{fig:SGDPoints}, while Figure \ref{fig:SGDObj} illustrates how the value of the objective function decreases over time.

 
\subsection{Nelder-Mead Method}

During the project, we also applied the Nelder-Mead method, a popular derivative-free method for unconstrained optimization, to minimize the penalized objective function \eqref{eq:PenaltyObj}. In particular, we utilize the MATLAB implementation of the Nelder-Mead method by calling the \texttt{fminsearch} function. Our experiments show that the method behaves fairly well.

\subsection{Coulomb Force Method}

All the methods we have considered so far are from an optimization perspective. In this section, we attempt to consider the problem using the original physical interpretation. According to Coulomb's law, electrons $i$ and $j$ repel each other with a force with magnitude proportional to $1/\|\bm{x}_i-\bm{x}_j\|^2$. In what we term as the \emph{Coulomb force method}, we first randomly arrange the $n$ points on the surface of the sphere, and then for each point, we simulate the effects of Coulomb's force that the other $n-1$ points exerted to this point by shifting it slightly in the corresponding direction. The summation of Coulomb's forces would clearly move the point out of the surface, and thus we project the point back to the surface of the sphere after each motion. We iterate this process through all the data points, and then carry out several passes through the data.

The intuition behind Coulomb's force method is that in each step we move a single point according to Coulomb's law while enforcing the spherical constraint; and then we iterate through all the data points several times in the hope that the process eventually stabilizes and we arrive at a state where the energy attains a local minimum. By implementing different initializations, there is also a chance that we might arrive at the global minimum. Clearly, the convergence of this method could not be guaranteed, yet our experiments show empirically that its performance is quite satisfactory, although its speed of convergence is rather slow as compared to previously discussed methods.

\section{Exploratory ideas and Future work}\label{sec:Explore}

\subsection{Convex Reformulation of the Constraint}

In this section, we consider a convex reformulation of the constraint $\|\bm{x}_j\|_2=1$. Inspired by \cite{PV2013}, we note that the following fact holds:
$$ \mathbb{E}\|\bm{A}\bm{x}_j\|_1 = \sum_{i=1}^m \mathbb{E}|\bm{a}_i^\mathsf{T}\bm{x}_j| = c\,m\|\bm{x}_j\|_2, $$
where $c = \sqrt{2/\pi}$ is a constant, $\bm{A}$ is a Gaussian random matrix of size $m\times k$ with its entries $a_{ij}\ i.i.d. \sim\mathcal{N}(0,1)$, and $\bm{a}_i$ denotes the $i$-th row of $\bm{A}$. The expectation is taken with respect to the elements of $\bm{A}$, and the last equation follows from the fact that the first absolute moment of the standard Normal distribution equals $c$.

The above observation indicates that the linear constraint $\|\bm{A}\bm{x}_j\|_1 = c\,m$ is equivalent to the original constraint $\|\bm{x}_j\|_2=1$  in expectation. Thus, when $n$ or $k$ becomes very large, instead of solving \eqref{eq:Thomson2} we can work with the following simpler problem instead:
\begin{equation}\label{eq:Thomson1}
\begin{aligned}
& \text{minimize}
& & \sum_{i=1}^n\sum_{j=1}^{i-1}\frac{1}{\|\bm{x}_i-\bm{x}_j\|_2^2} \\
& \text{subject to}
& & \bm{x}_i\in\mathbb{R}^k,\ \|\bm{A}\bm{x}_i\|_1 = c\,m,\ 1 \le i \le n.
\end{aligned}
\end{equation}

\subsection{The Sphere Packing Problem}

\begin{figure}[!htb]
\centering
\includegraphics[width=0.5\columnwidth]{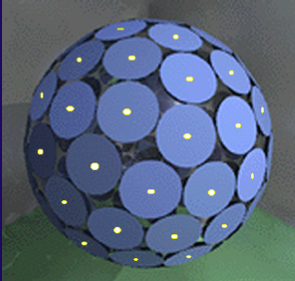}
\caption{The Sphere Packing Problem \cite{SpherePack}.}
\label{fig:SpherePack}
\end{figure}

We note that Thomson problem is closely related to the \emph{sphere packing problem}, which aims to find the largest diameter of \textit{n} identical circles that can be placed on the sphere without overlapping \cite{SpherePack}. Figure \ref{fig:SpherePack} provides an illustration of the problem.

A solution to the sphere packing problem should possess the following properties:
\begin{enumerate}[label=\arabic{*})]
	\item Each point $\bm{x}_i$ is equally distant to its three nearest neighbors.\label{Prop1}
	\item There exists $d$ such that $\min_{i \neq j} \| \bm{x}_i - \bm{x}_j \|= d$ for all $i$.
\end{enumerate}
According to \cite{SpherePack}, there is a simple algorithm for solving the problem:
\begin{enumerate}[label=\emph{Step \arabic{*}:}]
	\item Put each point in the centre of its three nearest neighbors as long as Property \ref{Prop1} is not satisfied.
	\item Calculate $d_i = \min_{i \neq j} \|\bm{x}_i - \bm{x}_j\|$ for all $i$. If the  $d_i$'s are not equal, take the point $\bm{x}_i$ with the largest value of $d_i$ and move its three nearest neighbors closer to $\bm{x}_i$. Then repeat Step 1.
	\item Use small random perturbations and search for the arrangement with greater minimum distance. When one is found, return to Step 1.
\end{enumerate}

\nocite{*}
\bibliography{Bibliography}

\begin{thebibliography}{1}

\bibitem{WikiThomson}
\url{http://en.wikipedia.org/wiki/Thomson_problem}.

\bibitem{WikiSphere}
\url{http://en.wikipedia.org/wiki/Spherical_coordinate_system}.

\bibitem{PV2013}
Y.~Plan and R.~Vershynin.
\newblock One-bit compressed sensing by linear programming.
\newblock {\em Communications on Pure and Applied Mathematics}, 66:1275--1297,
  2013.

\bibitem{SpherePack}
\url{http://www-lp.fmf.uni-lj.si/plestenjak/talks/preddvor.pdf}.

\bibitem{NW2006}
J.~Nocedal and S.~Wright.
\newblock {\em Numerical Optimization}.
\newblock Springer, 2nd edition, 2006.

\bibitem{Gleich2008}
D.~Gleich.
\newblock \url{https://www.cs.purdue.edu/homes/dgleich/cs520-2014/}.

\end{thebibliography}


\end{document}